\documentclass[a4paper,fleqn]{cas-sc}

\usepackage[numbers, sort&compress]{natbib}

\def\tsc#1{\csdef{#1}{\textsc{\lowercase{#1}}\xspace}}
\tsc{WGM}
\tsc{QE}

\begin{document}
\let\WriteBookmarks\relax
\def\floatpagepagefraction{1}
\def\textpagefraction{.001}

\shorttitle{ATFusion: An Alternate Cross-Attention Transformer Network for Infrared and Visible Image Fusion}    

\shortauthors{Han Yan et al.}  

\title [mode = title]{ATFusion: An Alternate Cross-Attention Transformer Network for Infrared and Visible Image Fusion}  


\tnotetext[1]{This work was supported in part by the National Natural Science Foundation of China (NSFC) under Grant 62101502, in part by China Postdoctoral Science Foundation under Grant 2022T150596, in part by the Key Research and Development and Promotion Foundation of Henan Province under Grant 232102211036, and in part by Postdoctoral Research Grant in Henan Province under Grant 202101012. } 

%

\author[1]{Han Yan}


\ead{han_yan0714@gs.zzu.edu.cn}


\credit{Conceptualization of this study, Methodology, Software}

\affiliation[1]{organization={Zhengzhou University},
            addressline={No.100, Kexuedadao Road}, 
            city={Zhengzhou},
            postcode={450001}, 
            state={Henan},
            country={China}}

\author[1]{Songlei Xiong}


\ead{xsl@gs.zzu.edu.cn}


\author[1]{Long Wang}


\ead{eielwang@gs.zzu.edu.cn}


\author[1]{Lihua Jian}
\ead{ielhjian@zzu.edu.cn}
\credit{supervision, writing}
\cormark[1]

\author[2,3]{Gemine Vivone}
\ead{gemine.vivone@imaa.cnr.it}

\affiliation[2]{organization={National Research Council, Institute of Methodologies for Environmental Analysis (CNR-IMAA)},
            city={Tito},
            postcode={85050}, 
            country={Italy}}

\affiliation[3]{organization={National Biodiversity Future Center(NBFC)},
            city={Palermo},
            postcode={90133}, 
            country={Italy}}

\cortext[cor1]{Corresponding author, Zhengzhou University, Zhengzhou, 450001, China}




\begin{abstract}
The fusion of infrared and visible images is essential in remote sensing applications, as it combines the thermal information of infrared images with the detailed texture of visible images for more accurate analysis in tasks like environmental monitoring, target detection, and disaster management. The current fusion methods based on Transformer techniques for infrared and visible (IV) images have exhibited promising performance. However, the attention mechanism of the previous Transformer-based methods was prone to extract common information from source images without considering the discrepancy information, which limited fusion performance. In this paper, by reevaluating the cross-attention mechanism, we propose an alternate Transformer fusion network (ATFusion) to fuse IV images. Our ATFusion consists of one discrepancy information injection module (DIIM) and two alternate common information injection modules (ACIIM). The DIIM is designed by modifying the vanilla cross-attention mechanism, which can promote the extraction of the discrepancy information of the source images. Meanwhile, the ACIIM is devised by alternately using the vanilla cross-attention mechanism, which can fully mine common information and integrate long dependencies. Moreover, the successful training of ATFusion is facilitated by a proposed segmented pixel loss function, which provides a good trade-off for texture detail and salient structure preservation. The qualitative and quantitative results on public datasets indicate our ATFusion is effective and superior compared to other state-of-the-art methods.
\end{abstract}


\begin{highlights}
\item Proposal of ATFusion Network: An end-to-end ATFusion network is introduced for infrared and visible (IV) image fusion, demonstrating superior performance and strong generalization across multiple datasets.

\item Dual Cross-Attention Feature Injection: A novel discrepancy information injection module (DIIM) and alternate common information injection module (ACIIM) are proposed to effectively extract and integrate both distinctive and shared features from source images.

\item Segmented Pixel Loss for Balanced Fusion: A segmented pixel loss function with intensity-specific constraints is designed to optimize texture detail preservation and brightness balance in the fused results.

\end{highlights}



\maketitle








\section{Introduction}
As the single-modality image has limited interpretation ability, it is difficult to meet the subsequent requirements for understanding real scenarios.
Therefore, developing an effective image fusion (IF) technology is urgently needed to assist people in gaining a deep understanding of images or tackling advanced computer vision tasks. Furthermore, image fusion (IF) can provide enhanced quality images for various applications, such as infrared target detection\cite{yuan2024c}, unmanned aerial vehicle (UAV) imaging\cite{mo2023robust}, and environmental monitoring\cite{zhang2022combined}. The resulting fused image merges thermal information with detailed texture, enhancing accuracy for remote sensing applications. Recent developments in VIF methods span traditional approaches\cite{ma2019infrared} to advanced deep learning techniques\cite{vivone2024deep,liu2018deep}.

A generally accepted classification for traditional fusion methods is as follows: multi-scale transform (MST)-based methods \cite{li2020mdlatlrr}, sparse representation (SR)-based methods \cite{mou2013image}, saliency-based methods \cite{jian2021infrared}, optimization-based methods \cite{ma2016infrared}, and hybrid-based methods \cite{hou2019infrared}. Among the traditional fusion methods, feature extraction and feature fusion are the two crucial steps, most of which are based on handcrafted techniques. However, blindly selecting complex transformations or representations for feature extraction often leads to time-consuming and information loss. Moreover, some manually designed fusion strategies may reduce final fusion performance as these fusion strategies are not optimized for the corresponding generated features.

Compared to traditional methods, recent advancements in deep learning have demonstrated superior computational efficiency and generalization. Existing proposed deep learning-based fusion methods are frequently implemented by leveraging either the convolutional neural network (CNN) techniques or Transformer techniques. Additionally, CNN-based methods can be further categorized into Autoencoder (AE)-based \cite{ren2021infrared}, end-to-end CNN-based \cite{xu2022multi,ding2021cmfa_net}, and GAN-based methods \cite{yi2021dfpgan}.

The AE-based approaches implement feature extraction and feature reconstruction through an encoder-decoder architecture, which needs to be trained in advance. However, as the handcrafted fusion strategies are not always compatible with deep extracted features, they inevitably hinder the improvement of fusion performance. Therefore, the end-to-end CNN-based approaches are introduced to directly generate the fused image under the specified constraints of a well-defined loss function. The GAN-based methods employ a min-max optimization game between generators and discriminators to solve the image fusion problem. Specifically, the generator is used to produce fusion results that closely match the distributions of the source image, and it can even deceive the discriminator. However, these CNN-based methods containing convolution operations can merely explore local information using fixed kernel size but lack consideration of the global information of an image.

By contrast, the Transformer is equipped with self-attention and cross-attention mechanisms, which can effectively extract long-range context information and has exhibited excellent performance in many vision tasks as the natural language processing \cite{chen2021vision}. Recently, Transformer-based architecture has been introduced into the image fusion community \cite{you2022hmf,tang2023datfuse}. These attempts have demonstrated long-range dependence, thanks to the self-attention and cross-attention mechanisms, further improving the quality of fusion outcomes \cite{ma2022swinfusion}.

Despite the comparable results achieved by existing Transformer-based fusion, several challenges still exist in this domain. First, the attention mechanism is only used to capture common information of source images, while the discrepancy information has not been effectively separated and utilized. Second, a single Transformer module cannot completely extract common information. The aforementioned issues significantly reduce fusion efficiency. Third, the existing pixel loss functions typically adopt a fixed approach, pixel maximum or weighted average, to guide the fusion process, which does not effectively preserve complete information.

To address the above-mentioned drawbacks, we devise an alternate Transformer fusion network (ATFusion) based on a modified cross-attention mechanism. For capturing the discrepancy information from two source images, we modify the cross-attention mechanism and propose a discrepancy information injection module (DIIM). To fully extract common information and integrate long dependencies of the source images, we design an alternate common information injection module (ACIIM). In addition, we propose a segmented pixel loss function that utilizes different constraints on pixel values to guide the fusion network. This proposed loss function can adequately ensure the preservation of intensity information in fused images.

The main contributions of this paper are summarized as follows:
\begin{enumerate}
    \item{An end-to-end ATFusion network is proposed for the fusion of IV images. Extensive experiments on multiple datasets show that our ATFusion method achieves good performance and generalization ability.}
    \item{A discrepancy information injection module (DIIM) and an alternate common information injection module (ACIIM) are proposed based on the cross-attention mechanism. With DIIM and ACIIM, the discrepancy and common features of the source images can be fully explored respectively.}
    \item{A segmented pixel loss function composed of different pixel intensity constraints is designed to train ATFusion so that the preservation of texture details and brightness information can reach a good trade-off in the fused results.}
\end{enumerate}

The rest of this paper is organized as follows. Section II provides a survey of the related work. Section III presents the proposed ATFusion framework. Section IV details the experimental analysis with some discussions. Finally, Section V summarizes the key findings of this paper.

\section{Related Work}
In this section, we review the current CNN-based and Transformer-based VIF methods.

\subsection{CNN-based Image Fusion Methods}
Deep learning (DL) has been widely introduced into computer vision tasks such as image restoration \cite{chen2020enhanced}, person re-identification \cite{yuan2018deep,ouyang2018video}, and image super-resolution \cite{zhang2018recent}. Attributed to the superior feature representation ability, DL-based VIF methods have also occupied the leading position in recent years. CNN was initially introduced into the field of image fusion by Liu \textit{et al.} \cite{liu2018infrared}, using Siamese convolutional networks to generate weight maps. Li \textit{et al.} \cite{li2018densefuse} proposed a dense-connection AE-based approach, DenseFuse, to fuse IV images, avoiding information loss of deep features in the fused results. Jian \textit{et al.} \cite{jian2020sedrfuse} designed an attention mechanism to fuse deep features based on a symmetric AE network, which enhanced the salient information of infrared images in the fused results. Li \textit{et al.} \cite{li2021rfn} trained a learnable fusion network at two stages to further preserve image details and overcome the limitations of hand-crafted fusion strategies. Similarly, Ma \textit{et al.} \cite{ma2021stdfusionnet} proposed an end-to-end model that utilized a salient target mask to guide the network training to highlight the thermal information. To improve the model generalization, Xu \textit{et al.} \cite{xu2020u2fusion} developed a unified unsupervised fusion framework to deal with different fusion tasks. To efficiently utilize the multi-scale deep features, Wang \textit{et al.} \cite{wang2021unfusion} proposed a multi-scale densely connected encoder-decoder fusion architecture. Moreover, Wang \textit{et al.} \cite{wang2022res2fusion} integrated the dense Res2Net and attention into the AE model to consider the long-range dependency features in the fusion results. 

Most GAN-based fusion methods force, in an unsupervised manner, the distribution of the fused result by a specific loss function. As a key milestone, Ma \textit{\textit{et al.}} \cite{ma2019fusiongan} first applied GAN to VIF. After that, various GAN-based variants have been proposed to boost the fusion performance. For example, Li \textit{et al.} \cite{li2019coupled} utilized the coupled generative adversarial network for the image fusion task. Ma \textit{et al.} \cite{ma2020ddcgan} designed a DDcGAN, in which the high- and low-versions of the fused image produced by a generator were used to deceive two discriminators respectively, to achieve the preservation of infrared information and detail information. To prevent the fused image from being biased to one of the source images, Ma \textit{et al.} \cite{ma2020ganmcc} adopted multiclassification constraints to achieve information balance. Wang \textit{et al.} \cite{wang2023cross} designed a cross-scale iterative attention manner to compute activity levels of different modal images based on the GAN model. To change the simple concatenation in generative adversarial fusion methods, Wang \textit{et al.} \cite{wang2022infrared} developed an interactive compensatory attention adversarial learning network.

\subsection{Transformer-based Fusion Methods}
Transformer was first proposed by Vaswani \textit{et al.} \cite{vaswani2017attention} to address natural language processing issues and achieved remarkable success. Subsequently, Dosovitskiy \textit{et al.} \cite{dosovitskiy2020image} designed visual Transformers to conduct an image classification task. Due to its self-attention mechanism that can capture long-range dependencies, the Transformer has been applied to many computer vision tasks, such as target detection \cite{xie2022focal,lin2022infrared}, video inpainting \cite{li2023feature,gan2022hybrid}, and image super-resolution \cite{zou2023multi,ariav2023fully}. Recently, some Transformer-based methods have been presented to handle VIF \cite{chen2023thfuse,yu2023end}.

Vs \textit{et al.} \cite{vs2022image} utilized a spatial branch and a Transformer branch fusion strategies to merge local and global information, respectively. Wang \textit{et al.} \cite{wang2022swinfuse} proposed a residual Swin Transformer to extract the global features that further enhanced the representation ability of the previous Transformer in VIF tasks. In a similar manner, Ma \textit{et al.} \cite{ma2022swinfusion} integrated cross-domain learning with Swin Transformer to fully use the inter- and intra-domain contexts for fusion. To overcome hand-crafted fusion rules existing in Transformer-based fusion models, Tang \textit{et al.} \cite{tang2023datfuse} proposed an end-to-end Transformer architecture to fuse IV images. 

Upon reviewing the previous Transformer-based fusion methods, we discovered certain limitations and rethought the usage of the cross-attention mechanism. Most studies based on Transformer structures merely grasp the global information of an image, aiming to compensate for the drawbacks of convolutional operations only capturing local information. To the best of our knowledge, both self-attention and cross-attention do not consider the discrepancy information between multimodalities when applied to VIF. Additionally, to fully exploit common information and long-range dependencies of source images, we adopt an alternating extraction approach depending on the original cross-attention mechanism.

\section{Methodology}
In this section, we first introduce the overall framework of the proposed ATFusion, discussing its core design principles. Next, we detail the modified cross-attention modules, DIIM and ACIIM, which are specifically tailored for image fusion tasks. Finally, we describe the loss function used to optimize the network.

\subsection{Framework Overview}

    \begin{figure*}[!t]
        \begin{center}
            \includegraphics[width=0.95\textwidth]{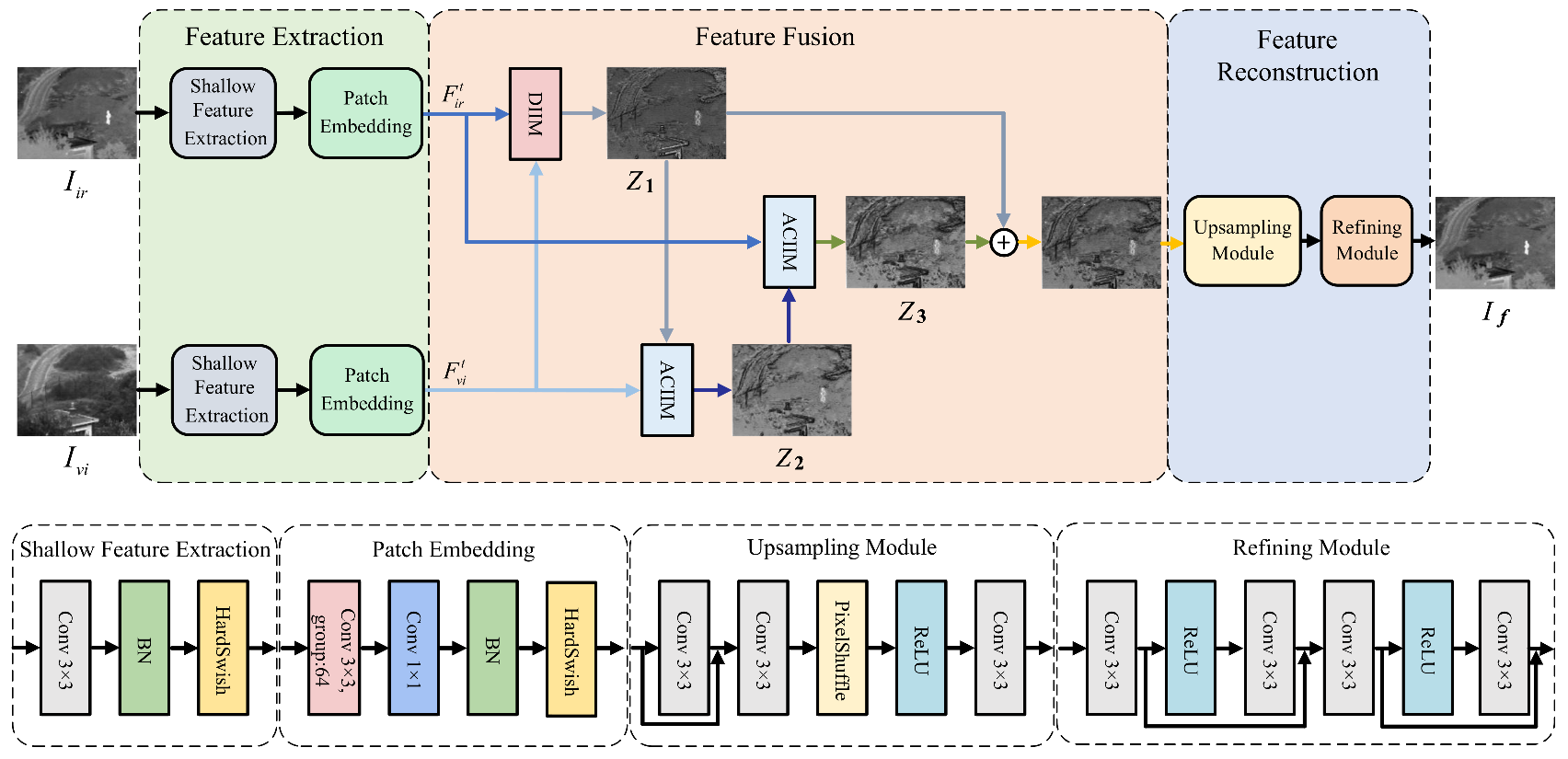}
        \end{center}
            \caption{The ATFusion framework: \textbf{Feature Extraction} for shallow feature and patch embedding, \textbf{Feature Fusion} via DIIM and ACIIM modules, and \textbf{Feature Reconstruction} with upsampling and refining to produce the fused image \( I_f \).}
            \label{framework}
    \end{figure*}

As illustrated in Fig.~\ref{framework}, the overall architecture of our ATFusion consists of a feature extraction module, a feature fusion module, and a feature reconstruction module. First, source IV images are fed into the feature extraction module to extract shallow features and then transform them into patch embedding. Afterward, a feature fusion module is established for fusing discrepancy features and common features. Finally, a feature reconstruction is utilized to map the fused features into a composite image. Here, we briefly introduce the proposed architecture.

\emph{Feature extraction}. Since convolutional layers have stability and improved optimization ability \cite{xiao2021early}, a convolutional layer with a kernel size of $3 \times 3$ is still used to extract shallow local features in the feature extraction module. After that, there is a batch normalization (BN) layer and a HardSwish activation function\cite{avenash2019semantic}. The feature extraction process can be formulated as:
\begin{equation}
    {\left\{F_{ir}^{sf}, F_{vi}^{sf}\right\}=\left\{SF(I_{ir}), SF(I_{vi})\right\}},
\end{equation}
where $I_{ir}$ and $I_{vi}$ are the source images, and $F_{ir}^{sf}$ and $F_{vi}^{sf}$ denote the output features of the shallow feature extraction, $SF(\cdot)$. Subsequently, we use the patch embedding module, $PE(\cdot)$, to transform the feature maps $F_{ir}^{sf}$ and $F_{vi}^{sf}$ to yield a series of token sequences $F_{ir}^{t}$ and $F_{vi}^{t}$, respectively. Then, these tokens are fed to the subsequent DIIM and ACIIM. The process of the patch embedding can be expressed as:
\begin{equation}
    {\left\{F_{ir}^{t}, F_{vi}^{t}\right\}=\left\{PE(F_{ir}^{sf}), PE(F_{vi}^{sf})\right\}.}
\end{equation}

\emph{Feature fusion}. The $F_{ir}^{t}$ and $F_{vi}^{t}$ are fed to the feature fusion module that is denoted as $FF(\cdot)$ to generate the fused feature $F_{f}$. The overall feature fusion process can be formulated as:
\begin{equation}
    {F_{f}=FF(F_{ir}^{t},F_{vi}^{t})},
\end{equation}
where the feature fusion module contains a DIIM ($DIIM(\cdot)$) and a pair of ACIIMs ($ACIIM(\cdot)$), which are designed to further extract global dependencies features, discrepancy features and common features, respectively. Specifically, the DIIM utilizes the query vectors provided by $F_{vi}^{t}$ and the key-value vectors provided by $F_{ir}^{t}$ to fuse the discrepancy features between the source images. It can be represented as:
\begin{equation}
    Z_{1} = DIIM(F_{ir}^{t},F_{vi}^{t}),\\  
\end{equation}
where $Z_{1}$ represents the fused discrepancy features that contain the main structures of the source images. To inject detail information into the $Z_{1}$ features, the first ACIIM uses the key-value vectors provided by $F_{vi}^{t}$ to explore the common features mainly derived from the visible image.     
\begin{equation}\label{aciim1}
    Z_{2} = ACIIM(F_{vi}^{t}, Z_{1}),\\  
\end{equation}
where $Z_{2}$ indicates the fused features enhanced by the detail information of the visible image. Similarly, the $Z_{2}$ features still need to be further enhanced by alternately injecting detail information from the infrared image using the second ACIIM as: 
\begin{equation}
    Z_{3} = ACIIM(F_{ir}^{t}, Z_{2}),\\ 
\end{equation}
where $Z_{3}$ are the fused features conveying the source images' discrepancy information and common information. The alternative detail injection may weaken the information about the main structure of the fused features. Hence, we again add the $Z_{1}$ features for preservation: 
\begin{equation}
    {F_{f}} = Z_{3} + Z_{1}.
\end{equation}
A detailed description of DIIM and ACIIM is provided in Section \ref{sec:DAA}.

\emph{Feature reconstruction}. The fused feature $F_{f}$ is fed into an upsampling module, $UP(\cdot)$, to resize back to the source images. Besides, we use the residual network-based refining module, $RE(\cdot)$, to further recover the detail information of the fused image, $I_{f}$. The feature reconstruction process can be expressed as:
\begin{equation}
    {I_{f}=RE(UP(F_{f})).}
\end{equation}

\subsection{DIIM and ACIIM}
\label{sec:DAA}
The goal of VIF is to obtain a comprehensive image that contains salient targets while preserving rich textural details. Thus, how to fully exploit discrepancy and common information present in the source images is a key determinant of the fusion performance. Motivated by the concept that the cross-attention mechanism is effective in exchanging information between two different patches \cite{2021CrossViT}, we propose DIIM and ACIIM to fuse discrepancy and common features among the source images.

\begin{figure}[!t]
    \begin{center}
        \includegraphics[width=0.45\textwidth]{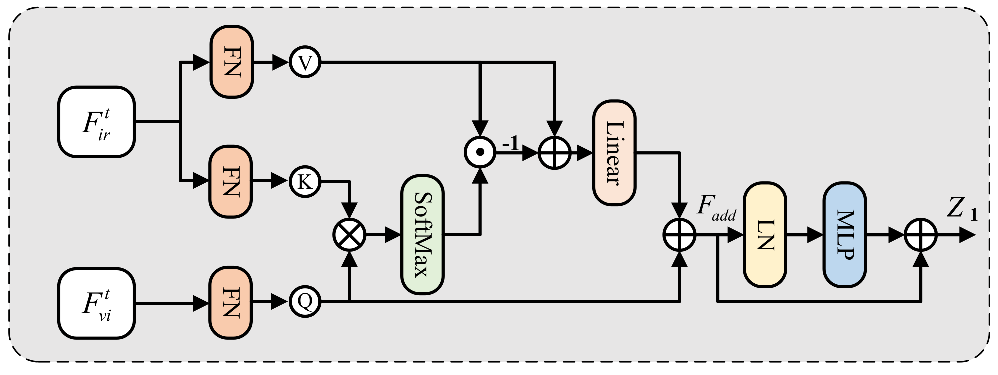}
    \end{center}
        \caption{The architecture of the discrepancy information injection module (DIIM). }
        \label{DIIMfig}
\end{figure}

To effectively obtain the discrepancy between the infrared and visible image features produced by the previous stage, we use a newly built cross-attention architecture, as illustrated in Fig. \ref{DIIMfig}. It is given the $F_{ir}^{t}$ and $F_{vi}^{t}$ as input and the discrepancy information features as output. The $FN$ module consists of two operation steps. First, to explore the long-range relationship of IV features, we partition, through a partition operator $\mathcal{P}$, the $F_{ir}^{t}$ and $F_{vi}^{t}$ into $s$ local feature segments as:
\begin{equation}
    \begin{aligned}
    &Q_1, \dots, Q_s=\mathcal{P}(F_{vi}^{t}),\\
    &K_1, \dots, K_s=\mathcal{P}(F_{ir}^{t}),\\
    &V_1, \dots, V_s=\mathcal{P}(F_{ir}^{t}),\\
    \end{aligned}
\end{equation}
where $F_{vi}^{t} \in \mathbb{R}^{h \times w \times c}$, $F_{ir}^{t} \in \mathbb{R}^{h \times w \times c}$, and $s= h\times w$. Afterward, we employ the linear layer to transform the token segments into the query vectors ($Q$), key vectors ($K$), and value vectors ($V$), which are three fundamental elements of the base Transformer. The linear projection can be expressed as:
\begin{equation}
    \begin{aligned}
    &Q_i=\text{Linear}_Q(Q_i),\\
    &K_i=\text{Linear}_K(K_i),\\
    &V_i=\text{Linear}_V(V_i),\\
    \end{aligned}
\end{equation}
where $i = 1, \dots, s$ and $\text{Linear}_{\cdot}(\cdot)$ are linear projection operators.

To explore the common information $CM_{inf}$ of infrared and visible image features with the consideration of long-term relationships, we use the dot-production attention layer to compute the similarity matrix between $Q_i$ and $K_j$ ($i$ and $j$ from 1 to $s$), and then multiply by the vector $V$ to infer the common information. This process can be expressed as:
\begin{equation}
    \begin{aligned}
    CM_{inf} = \text{SoftMax}\left(\frac{Q_{1,\dots,s} K^T_{1,\dots,s}}{\sqrt{d_k}}\right) V ,
    \end{aligned}
\end{equation}
where $\cdot^T$ is the transpose operator and $d_k$ is a scaling factor, which can alleviate the softmax function, $\text{SoftMax}\left(\cdot\right)$, from converging to regions of minimal gradients as the dot product increases. Subsequently, we can easily obtain the discrepancy information, $DIM_{inf}$, between the infrared and visible images by removing the common information. This process can be represented as:
\begin{equation}
    \begin{aligned}
    & DIM_{inf} = \text{Linear}(V - CM_{inf}).
    \end{aligned}
\end{equation}
    
To obtain complementary information, $F_{add}$, from the IV images, we inject the discrepancy information into $Q$, which can be formulated as: 
\begin{equation}
    F_{add} =  DIM_{inf} + Q.
\end{equation}
    
Then, the final fused discrepancy features are obtained by layer normalization, $LN(\cdot)$, nonlinear mapping of the $F_{add}$ features, and skip connection operation:
\begin{equation}
    Z_{1} = MLP(LN(F_{add})) + F_{add},
\end{equation}
where $MLP(\cdot)$ includes two $\text{Linear}(\cdot)$ and one activation function $GELU(\cdot)$, and $Z_{1}$ represents the output of the DIIM.

\begin{figure*}[!t]
    \begin{center}
        \includegraphics[width=0.95\textwidth]{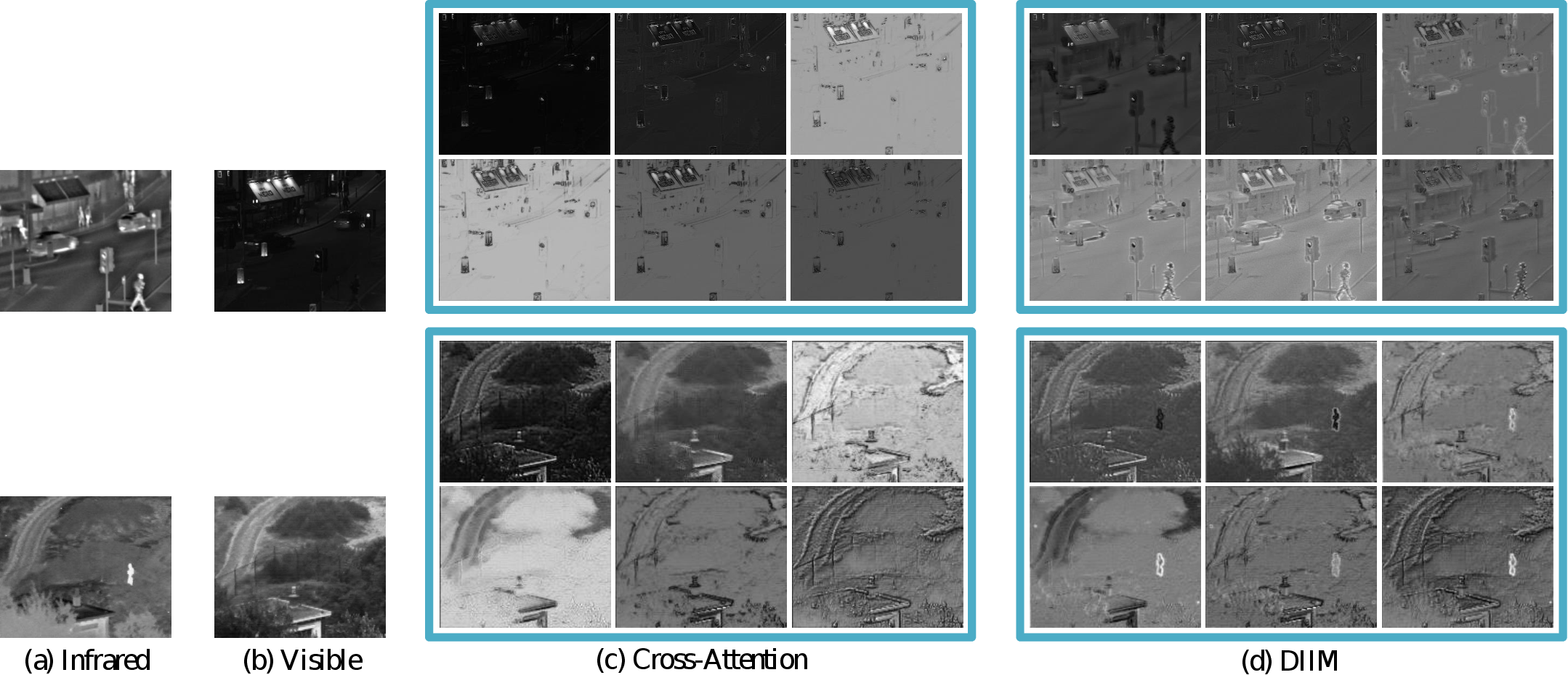}
    \end{center}
        \caption{Partial feature maps obtained from the two different modules. (a) infrared image, (b) visible image, (c) partial features obtained by the vanilla cross-attention, (d) partial features obtained by the DIIM.}
        \label{Featurefig}
\end{figure*}

To illustrate the effectiveness of the modified cross-attention, we compare the feature maps generated by the DIIM with those generated by the vanilla cross-attention. The results are shown in Fig.~\ref{Featurefig}. The feature maps illustrate that the information injection module utilizing the vanilla cross-attention mechanism can solely incorporate common information from both images, lacking the capability to integrate discrepancy information across different modalities. Consequently, the resultant output feature maps predominantly encompass information from one image, while lacking modality-specific details from the other image, making them unsuitable for multimodal image fusion tasks. In contrast, our DIIM significantly compensates for this deficiency.

\begin{figure}[!t]
    \begin{center}
        \includegraphics[width=0.45\textwidth]{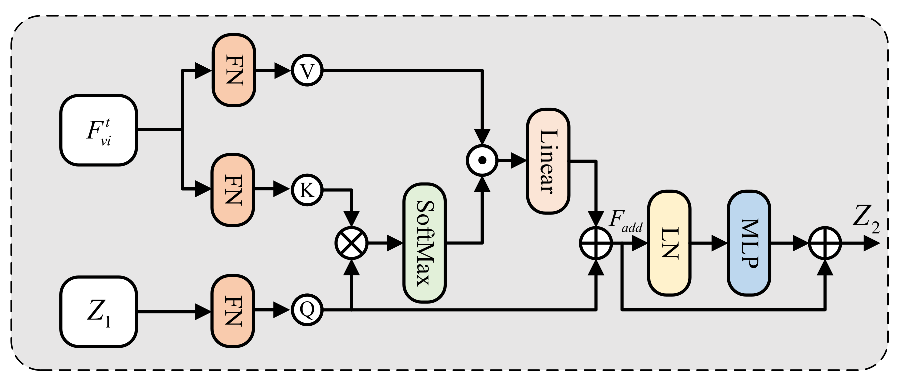}
    \end{center}
        \caption{The architecture of the alternate common information injection module (ACIIM).}
        \label{ACIIMfig}
\end{figure}

As depicted in Fig.~\ref{Featurefig}, although the obtained feature maps generated by the DIIM almost contain all salient edge information, the background detail information is not completely preserved in the fusion features. Thus, it is essential to further enhance the fused features by incorporating common information from both IV images. Following the DIIM, the proposed ATFusion alternately extracts common information reflecting the background details from the IV images through the ACIIM module. The structure of the ACIIM is illustrated in Fig.~\ref{ACIIMfig}. We expand Eq.~(\ref{aciim1}). First, given that the segments of $Z_{1}$ are used as $Q_{1, \dots , s}$ and the segments of $F_{vi}^{t}$ are used as $K_{i, \dots, s}$ and $V$, the common information, $CM_{vi}^{t}$, between $Z_{1}$ and $F_{vi}^{t}$ can be expressed as:
\begin{equation}\label{eq:common_vip}
    \begin{aligned}
    CM_{vi}^{t} = \text{SoftMax}\left(\frac{Q_{1,\dots,s} K^T_{1,\dots,s}}{\sqrt{d_k}}\right) V.
    \end{aligned}
\end{equation}

Then, $CM_{vi}^{t}$ is added to $Z_{1}$ to compensate for the details of the fused discrepancy features. The process can be formulated as: 
\begin{equation}\label{eq:enhance}
    \begin{aligned}
    F_{add} &= \text{Linear}(CM_{vi}^{t}) + Q, \\
    Z_{2} &= MLP(LN(F_{add})) + F_{add},
    \end{aligned}
\end{equation}
where $Z_{2}$ represents the output of the first ACIIM.

Similarly, we compute the common information between $Z_{2}$ and $F_{ir}^{t}$ to extract the detail information from the infrared image according to Eqs.~(\ref{eq:common_vip}) and~(\ref{eq:enhance}).

\subsection{Loss Function}
\label{sec:Loss}
Since the proposed ATFusion is trained in an unsupervised end-to-end manner, the choice of a loss function greatly affects the fusion performance. Considering the different imaging mechanisms of IV images, the loss function of the proposed method must ensure that the fusion results can preserve sufficient details and salient information. The loss function of our ATFusion can be defined as:
\begin{equation} \mathcal{L} = \mathcal{L}_{pixel} + \gamma \cdot \mathcal{L}_{texture}, \end{equation}
where $\mathcal{L}_{pixel}$ and $\mathcal{L}_{texture}$ represent the pixel loss and texture loss, respectively. $\gamma$ is a hyperparameter to balance these two loss terms. 

Motivated by \cite{tang2022piafusion}, the texture details of an image can be represented by maximum aggregation around the gradients. Thus, the texture loss is designed to regulate the gradients in the fused image as follows:
\begin{equation}
    \mathcal{L}_{texture}=\frac{1}{H \times W} \Vert \nabla I_{f}-max\{\nabla I_{ir}, \nabla I_{vi}\}  \Vert_1,
\end{equation}
where $\nabla$ represents the Sobel operator, which is used to calculate the gradient, $\Vert \cdot \Vert_1$ represents the  $\ell_{1}$-norm, $H$ and $W$ denote the height and the width of the image, respectively, and $max\{\cdot,\cdot\}$ indicates the element-wise maximum operator.

It is known that the maximum value selection-based pixel loss weakens the importance of pixels in one of the source images, while the average pixel loss may reduce the saliency of fusion results. Therefore, to achieve a better compromise between the preservation of important information and the enhancement of salient information, we adopt a segmented pixel loss function to train the proposed framework. In this work, the importance of a pixel, $\hat{p}$, is defined as the product of the pixel value and its gradient value:
\begin{equation}
    \hat{p}_{i,j}=\nabla I_{i,j} \cdot I_{i,j},
\end{equation}
where $ I_{i,j} $ represents the pixel value of the image $ I $ at the $ (i, j) $ position, while $ \nabla I_{i,j} $ denotes the gradient value at the corresponding position. In this context, $ \hat{P} $ and $ \hat{Q} $ represent the pixel importance matrices of the infrared and visible light images, respectively.

The importance of each pixel in the two source images is assessed and categorized into two segments as follows:
\begin{equation}\label{segmentseq}
    \begin{aligned}
        & \hat{P}_{top} = \{(i,j) \mid \hat{P}_{i,j} \geq \hat{p}_{ir}^{\alpha} \ \text{or} \ \hat{Q}_{i,j} \geq \hat{q}_{vi}^{\alpha}\} \\
        & \hat{P}_{res} = \{(i,j) \mid \hat{P}_{i,j} < \hat{p}_{ir}^{\alpha} \ \text{and} \ \hat{Q}_{i,j} < \hat{q}_{vi}^{\alpha}\}
    \end{aligned}
\end{equation}
where $\hat{p}_{ir}^{\alpha}$ and $\hat{q}_{vi}^{\alpha}$ represent the top $\alpha\%$ important pixels, and $\hat{P}_{top}$ denotes the first segment that consists of the most crucial pixels covering the top $\alpha\%$ important pixels from each source image. The remaining pixels are represented as the second segment, $\hat{P}_{res}$.

As depicted above, we apply different pixel loss functions to each segment to meet the trade-off among pixels of different importance. The maximum value selection-based pixel loss is employed in $\hat{P}_{top}$, which can emphasize the saliency of these pixels in the fused image. Meanwhile, for $\hat{P}_{res}$, we use the pixel average loss to force the fused image to approximate each source image. The segmented pixel loss can be formulated as follows:
\begin{equation}
    \begin{aligned}
        \mathcal{L}_{top} = &\frac{1}{N_{top}} \sum_{(i,j) \in \hat{P}_{top}}\Vert I_f(i,j) - max(I_{ir}(i,j),I_{vi}(i,j)) \Vert ,\\
        \mathcal{L}_{res} = & \frac{1}{2N_{res}} \sum_{(i,j) \in \hat{P}_{res}} \left( \Vert I_f(i,j) - I_{ir}(i,j) \Vert \right) \\
        &       + \frac{1}{2N_{res}} \sum_{(i,j) \in \hat{P}_{res}} \left( \Vert I_f(i,j) - I_{vi}(i,j) \Vert \right)
    \end{aligned}
\end{equation}

where \( \mathcal{L}_{top} \) and \( \mathcal{L}_{res} \) are the loss functions associated with the top \( \alpha\% \) most important pixels and the remaining pixels, respectively. The number of pixels in \( \hat{P}_{top} \), denoted as \( N_{top} \), is given by \( N_{top} = H \times W \times \alpha\% \), and the number of pixels in \( \hat{P}_{res} \), denoted as \( N_{res} \), is \( N_{res} = H \times W \times (1 - \alpha\%) \). The final pixel-wise loss function, \( \mathcal{L}_{pixel} \), is defined as:

\begin{equation}
    \begin{aligned}
        &\mathcal{L}_{pixel} = \mathcal{L}_{top} + \mathcal{L}_{res}.
    \end{aligned}
\end{equation}

\section{Experiments}
In this section, we first introduce the datasets and implementation settings. Then, the comparative experimental methods and objective evaluation metrics are described. After that, the experiment results and discussion are reported. Furthermore, we present the ablation experiments and the analysis of some hyperparameters and computational efficiency. 

\subsection{Datasets and Implementation Settings}
In this study, we use three public datasets for our experiments: MSRS \cite{tang2022piafusion}, RoadScene \cite{xu2020u2fusion}, and RGB-NIR Scene \cite{brown2011multi}, see Tab.~\ref{data}. The raw MSRS dataset contains 1,444 pairs of aligned IV images (training: 1083; test: 361). We randomly split the training dataset into 900 pairs for training and 180 pairs for validation. The training and validation image pairs are then randomly cropped in sample patches with a size of $128\times128$. All test pairs in the MSRS dataset are used to verify the superior performance of the proposed method. To demonstrate the generalization ability, we chose 20 image pairs from the RoadScene dataset and 20 typical image pairs from the RGB-NIR Scene dataset.

	\begin{table}[t]
		\caption{Datasets used for experiments (unit: image pairs)}
		\begin{center}
			\begin{tabular}{p{2.3cm}p{1.3cm}p{1.3cm} p{1.3cm}}
				\hline
				Dataset &Training &Validation &Test\\
				\hline
				MSRS &900 &180 &361\\
				\cline{2-4} 
				RoadScene &--- & ---&20\\
                \cline{2-4}
                RGB-NIR Scene &--- &--- &20  \\
				\hline					
			\end{tabular}
			\label{data}
		\end{center}
	\end{table}

To train the proposed ATFusion, we use the AdamW optimizer to update network parameters with an initial learning rate of $1\times 10^{-4}$. The learning rate drops to half of the previous one at 50, 100, 200, and 400 epochs. The batch size is set to 16. The parameters $\alpha$ and $\gamma$ are set to 20 and 1.0, respectively. Data augmentation techniques, including random horizontal flipping, noise injection, random cropping, and brightness adjustment, are applied to the original images with a probability of $20\%$. Additionally, a weight regularization constraint is incorporated to improve the model's generalization and stability. The configurations for all experiments are computed with an NVIDIA GeForce RTX 4090 GPU and 64 GB RAM. The framework is programmed in PyTorch.

\subsection{Comparative Methods and Evaluation Metrics}

\emph{Comparative Methods}: To assess the superiority of our method, we select eight state-of-the-art image fusion methods for comparison, including DenseFuse \cite{li2018densefuse}, U2Fusion \cite{xu2020u2fusion}, RFN-Nest \cite{li2021rfn}, SwinFuse \cite{wang2022swinfuse}, SwinFusion \cite{ma2022swinfusion}, LRRNet \cite{li2023lrrnet}, DATFuse \cite{tang2023datfuse}, and AEFusion \cite{li2023aefusion}. The source codes of all compared methods are publicly available and all parameters are configured as suggested in the original papers. For a fair comparison, these state-of-the-art methods are trained on the same dataset to obtain fusion models, and the fusion results are generated on the same test datasets.  

\emph{Evaluation Metrics}: Eight objective evaluation metrics are chosen to quantitatively evaluate the performance of all methods: average gradient (AG), entropy (EN) \cite{liu2015general}, feature mutual information (FMI) \cite{haghighat2011non}, standard deviation (SD) \cite{rao1997fibre}, spatial frequency (SF) \cite{zheng2007new}, structural similarity (SSIM) \cite{wang2004image}, visual information fidelity (VIFF) \cite{han2013new}, and $Q_{abf}$ \cite{piella2003new}. AG measures the sharpness or clarity of an image. EN is an objective measure of the amount of information contained in an image. FMI evaluates the total of feature information transferred to the fused image from the source images. $Q_{abf}$ is a full-reference quality evaluation index estimating the degree of salient information from source images represented in the fused image. SD is a statistical theory-based standard deviation that reflects the degree of change in pixel brightness. SF is a measure based on gradient distribution, which represents the rate of change of the grayscale of the fused image. SSIM measures the closeness between the fused image and the source images from the brightness, structure, and contrast aspects. VIFF measures the quality of the fusion image according to the visual information fidelity. Higher values of these metrics correspond to better quality of the fused image.

\subsection{Results and Discussion}

\begin{figure*}[!t]
    \begin{center}
        \includegraphics[width=0.6\textwidth]{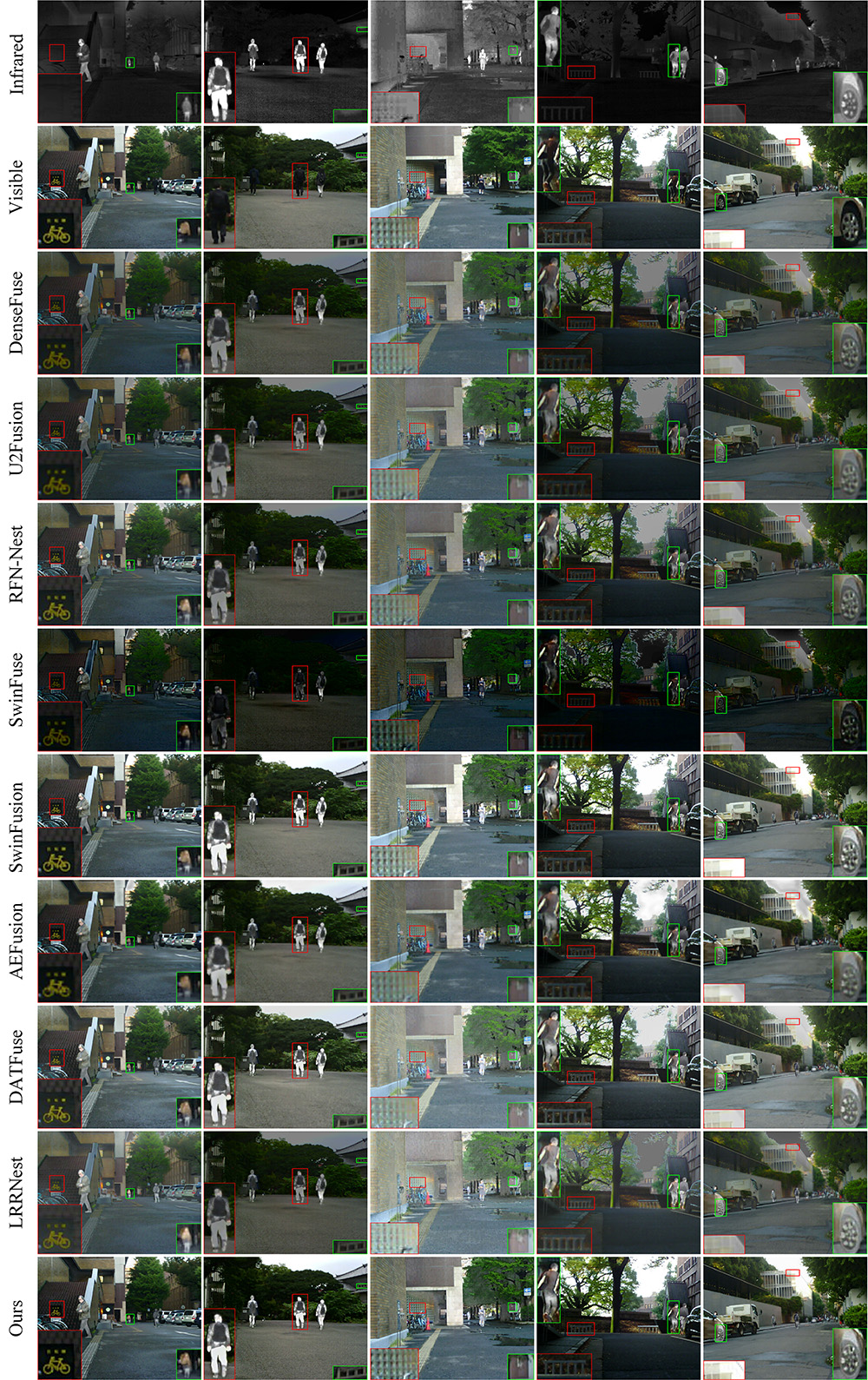}
    \end{center}
    	\caption{Qualitative assessment of the compared VIF methods on the MSRS dataset. }
        \label{msrs}
\end{figure*}

\emph{1) Results on the MSRS Dataset}: Fig.~\ref{msrs} displays the fusion results on the MSRS dataset. Although the existing fusion methods can obtain results by retaining information from the source images to some extent, they still encounter detail blurring and salient information loss. Specifically, DenseFuse, SwinFuse, RFN-Nest, U2Fusion, and LRRNet fail to preserve well brightness information from the infrared image in the fusion results (see the red box in the second column and the green box in the fourth column). The results generated by AEFusion lose different degrees of detail information, causing blurriness in some areas, see the red box in the third column and the green box in the fifth column. The results of DATFuse also exhibit poor performance in terms of texture information injection of the visible image and thermal target preservation of the infrared image, see the green box in the third column and the red box in the fourth column. SwinFusion seems to have richer details and brightness information. However, it does not retain some details from the infrared images, since this method utilizes a pixel maximum constraint loss for each pixel, see the red boxes in the fourth and fifth columns. In contrast, our ATFusion achieves a better trade-off between detail preservation and salient information retention.

\begin{table*}[!htbp]\scriptsize
    \caption{Quantitative assessment of the compared approaches on the MSRS dataset (361 image pairs). The best and the second best performance is marked by \textbf{bold} and \underline{underlined}, respectively.}
		\label{tab1}
		\begin{center}
			\begin{tabular}{ l l l l l l l l l l l l l}
				\hline
				Methods       &AG  &EN  &FMI  &$Q_{abf}$ &SD  &SF &SSIM   &VIFF     \\
				\hline	
DenseFuse      &2.0441   &5.9331   &0.9232   &0.3651   &7.4304  &0.0235  &0.9300  &0.7053 \\

U2Fusion      &2.0000   &5.8656 	&0.9134  &0.3161  &7.4645  &0.0226  &0.9120  &0.6487  \\

RFN-Nest    &2.5485   &6.2196 	&0.9229  &0.5042  &7.7855  &0.0296  &0.9554  &0.7828  \\

SwinFuse   &1.9669   &4.9449 	&0.8527  &0.2038  &5.7161  &0.0262  &0.9052  &0.5980  \\

SwinFusion   &\underline{3.5414}   &\textbf{6.6179}	&\underline{0.9312}  &\underline{0.6198}  &\underline{8.3973}  &\underline{0.0434}  &\underline{0.9695}  &\underline{1.0039}  \\

AEFusion    &2.7009   &6.5641 	&0.9088  &0.4319  &8.2501  &0.0305  &0.9293  &0.8662  \\

DATFuse    &3.5227   &\underline{6.5817} 	&0.9085  &0.5709  &\textbf{8.7157}  &0.0413  &0.9394  &0.7980  \\

LRRNet        &2.6309   &6.3140 	&0.9097  &0.3539  &8.2313  &0.0293  &0.8872  &0.7397  \\

Ours     &\textbf{3.7619}   &6.5784 	&\textbf{0.9318}  &\textbf{0.7022}  &8.2459  &\textbf{0.0453}  &\textbf{0.9759}  &\textbf{1.0370}  \\
        \hline
            \end{tabular}
        \end{center}
        \label{tabMSRS}
\end{table*}

Tab.~\ref{tabMSRS} displays the quantitative assessment of the compared approaches on 361 image pairs from the MSRS dataset. Our ATFusion method performs best in most metrics, except for EN and SD. In the EN case, it ranks third, scoring slightly less than the best. The reason for the relatively low value of the SD metric is that the DIIM module is used in our method, which makes the pixel distribution of the fused image relatively peaked. Overall, based on both the subjective and objective analyses, we can conclude that our ATFusion method achieves the best fusion performance and outperforms the competitors.

\begin{figure*}[!t]
    \begin{center}
        \includegraphics[width=0.6\textwidth]{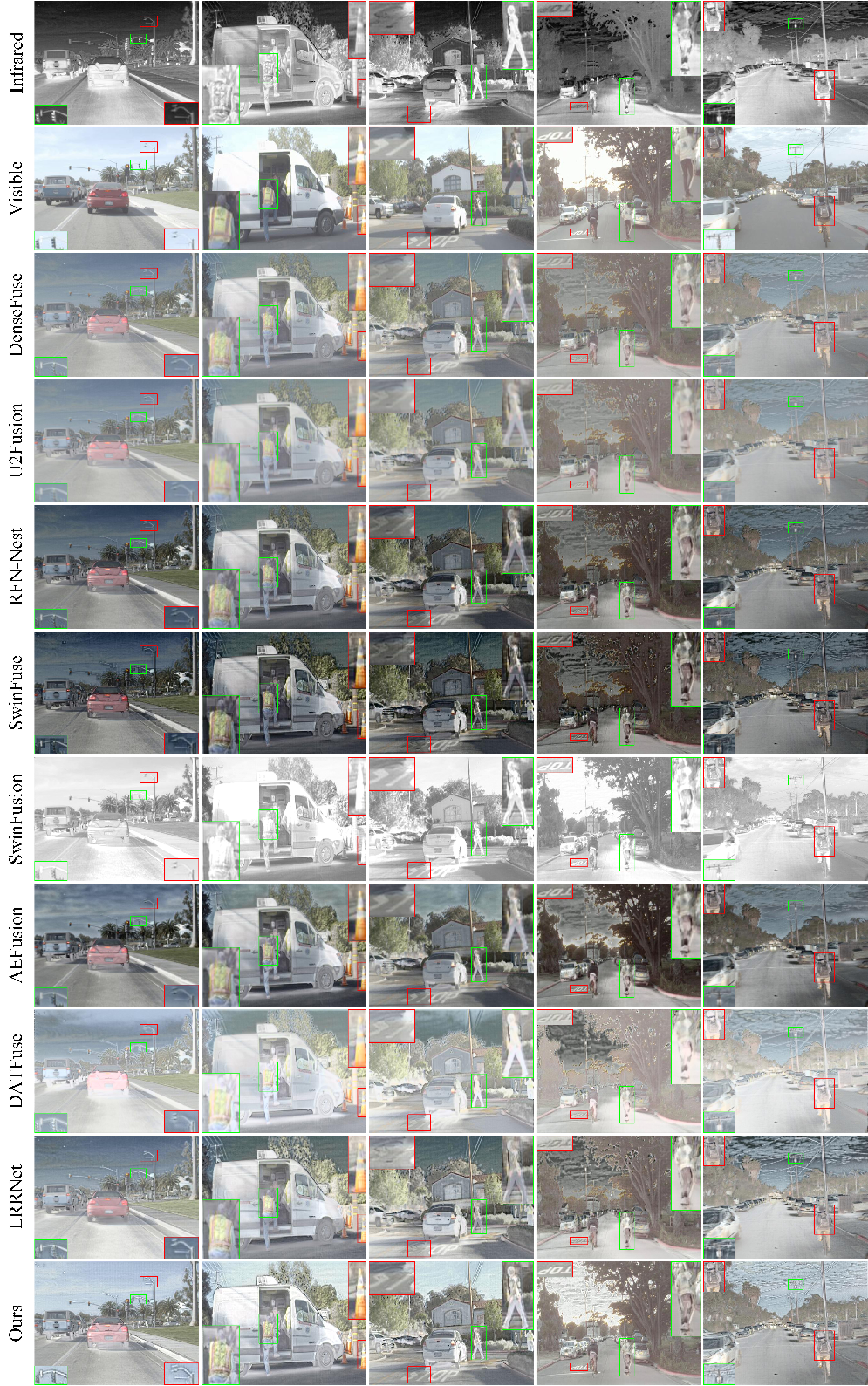}
    \end{center}
    \caption{Qualitative assessment of the compared approaches on the RoadScene dataset.}
    \label{roadScene}
\end{figure*}

\emph{2) Results on the RoadScene Dataset}: Fig.~\ref{roadScene} shows five pairs of source infrared and visible images from the RoadScene dataset along with the fusion results obtained through the compared methods. Two localized areas in each image are magnified for better comparison. It is clear that the state-of-the-art methods used for comparison have some unsatisfactory performance. Indeed, results generated by DATFuse, AEFusion, U2Fusion, Densefuse, and SwinFusion have deficiencies when dealing with specific local information, see the red box in the first column and the green box in the second column. Besides, RFN-Nest fails to preserve well significant information from the infrared image in the fusion results, see the red and green boxes in the fifth columns. Although the fusion results of LRRNet and SwinFuse can highlight well the target information in the source image, they fail to transfer the detail information to the fused images in some scenes, see the red box in the third and fourth columns. By contrast, it can be seen that our ATFusion maintains the salient information in all results. This is because the proposed ATFusion employs the DIIM and ACIIM modules, which can generate results containing various types of information. Furthermore, the segmented pixel loss function enables our method to better balance the retention of detailed and salient information.

\begin{figure*}[!t]
    \begin{center}
        \includegraphics[width=0.95\textwidth]{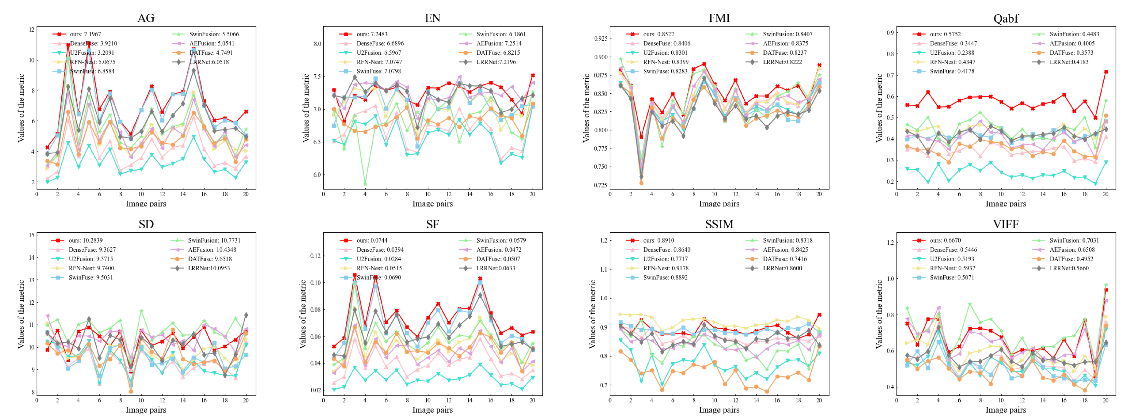}
    \end{center}
    \caption{Quantitative assessment of the compared approaches on the RoadScene dataset (20 image pairs). The average for all metrics is shown in the legend.}
    \label{roadScenceMetric}
\end{figure*}

Fig.~\ref{roadScenceMetric} further reports quantitative results on the 20 image pairs of the RoadScene dataset by using the eight quality metrics. These results indicate that our method gets the best average values on four evaluation metrics (i.e., AG, SF, $Q_{abf}$, and FMI). In the case of the EN, VIFF, and SSIM quality metrics, it ranks second. Overall, our method achieves satisfactory objective performance, which demonstrates the crucial role of the proposed framework and the segmented pixel loss function.

\begin{figure*}[!t]
    \begin{center}
        \includegraphics[width=0.6\textwidth]{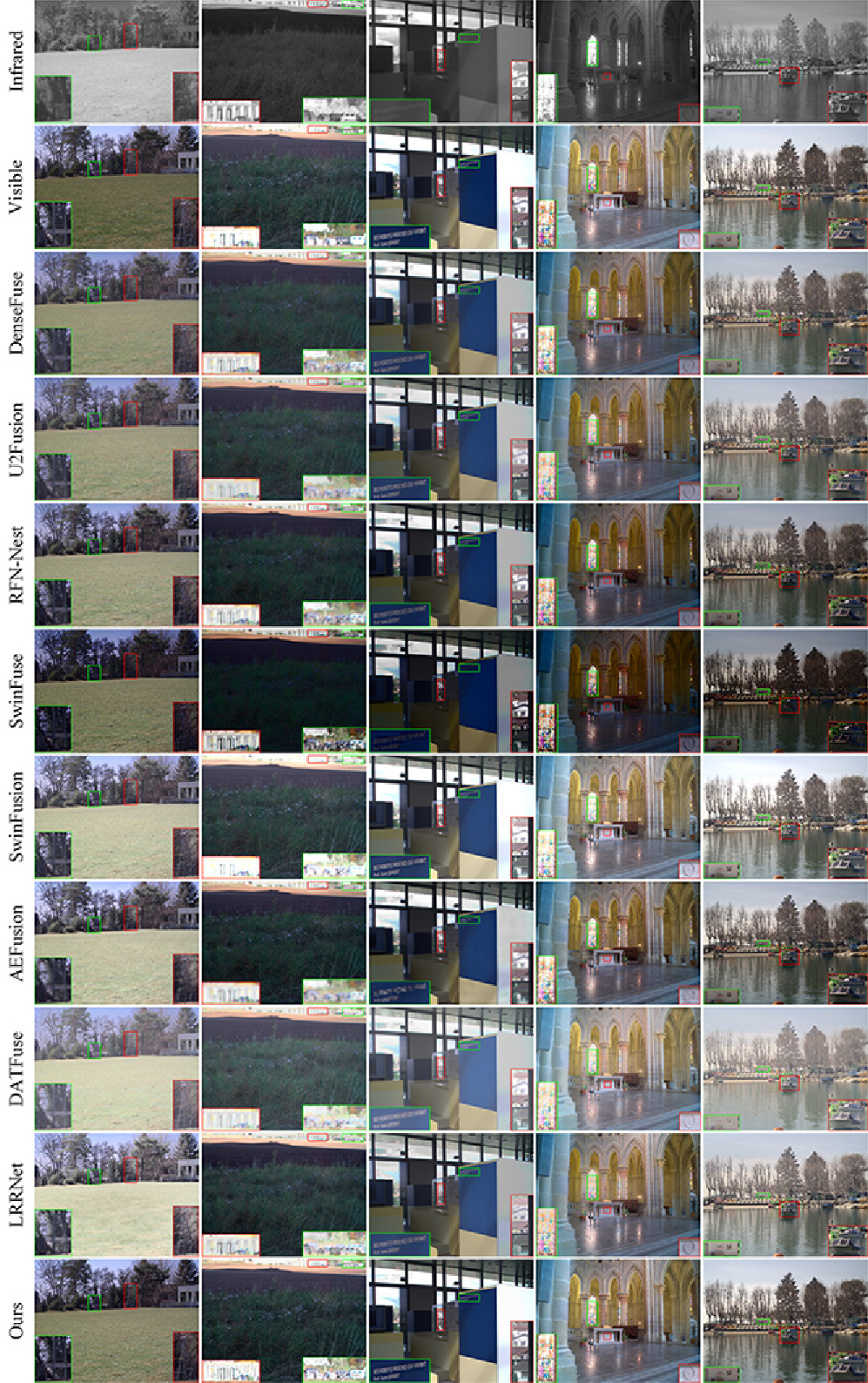}
    \end{center}
    	\caption{Qualitative assessment of the compared approaches on the RGB-NIR Scene dataset.}
        \label{nirscene}
\end{figure*}

\emph{3) Results on the RGB-NIR Scene Dataset}: To further validate the effectiveness and generalization capabilities of our method, we directly utilize the trained model to test the RGB-NIR Scene dataset. Fig.~\ref{nirscene} shows five image pairs from the RGB-NIR Scene dataset and the corresponding fusion results obtained by nine methods. We have magnified two local areas in each image for better visual comparison. It can be seen that the compared state-of-the-art methods still have some drawbacks compared to our approach. Specifically, SwinFusion, DATFuse, AEFusion, RFN-Nest, U2Fusion, and LRRNet fail to extract enough salient information from the infrared images compared to the other methods, see the green and red boxes in the second column. Moreover, the results produced by DenseFuse and SwinFuse are blurry in some local details, see the green box in the first and third columns. Compared with these methods, ATFusion cannot only preserve more detail, but also show a higher degree of target salience.

\begin{figure*}[!t]
    \begin{center}
        \includegraphics[width=0.95\textwidth]{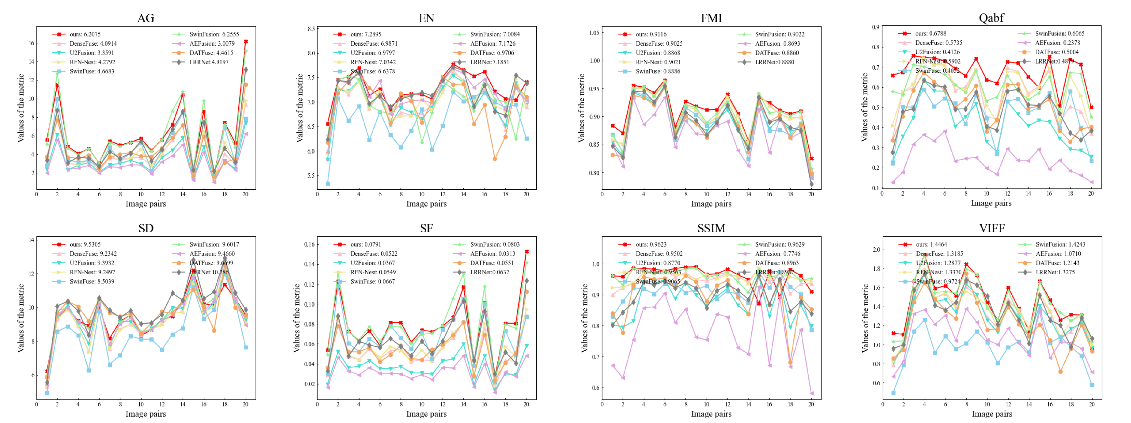}
    \end{center}
    \caption{Quantitative assessment of the compared approaches on the RGB-NIR Scene dataset (20 image pairs). The average for all metrics is shown in the legend. }
    \label{RGB-NIR Scene Metric}
\end{figure*}

Fig.~\ref{RGB-NIR Scene Metric} reports that our method achieves the best fusion performance for the EN, VIFF, $Q_{abf}$, and FMI metrics on the RGB-NIR Scene dataset. Instead, it is the second-best for the AG, SSIM, and SF metrics, with just a small gap compared to the best performance. A possible reason for the low SD value is that the pixels of the generated image are concentrated around the mean by using the DIIM module. In general, the ATFusion method achieves the best objective and subjective performance among the compared approaches on the 20 image pairs from the RGB-NIR Scene dataset, corroborating its excellent generalization ability.

\subsection{Ablation Studies}
In this section, we have conducted two ablation studies, i.e., considering the network structure and the segmented pixel loss function.
    
\begin{figure*}[!t]
    \begin{center}
        \includegraphics[width=0.95\textwidth]{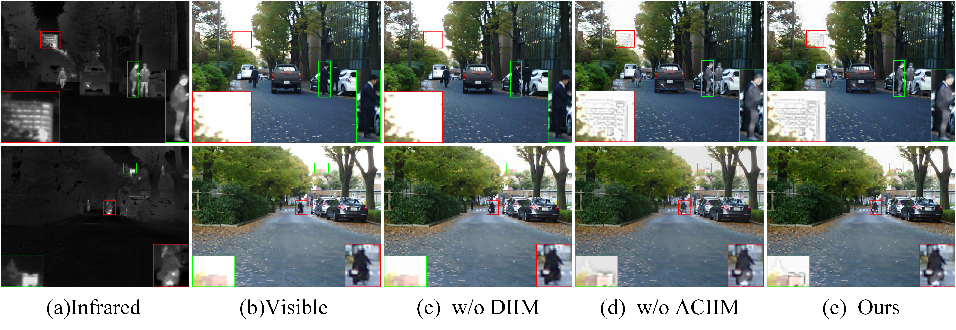}
    \end{center}
    	\caption{Fusion results obtained by three different network structures. (a) Infrared image, (b) visible image, (c) the results w/o DIIM, (d) the results w/o ACIIM, (e) Ours. }
        \label{fig14}
\end{figure*}

\emph{1) Ablation Study on the Network Structure}: To investigate the effectiveness of DIIM and ACIIM, we separately remove the DIIM (named ``w/o DIIM"), and the ACIIM (named ``w/o ACIIM"). We compare the fusion results obtained by w/o ACIIM, w/o DIIM, and the proposed method, as shown in Fig.~\ref{fig14}. It can be seen that the fusion results produced by using w/o DIIM only contain detail information from the visible image, while the salient information from the infrared image is almost entirely lost, see the green and red boxes in Fig.~\ref{fig14}(c). Besides, our fusion results have more detail information than the results by using the w/o ACIIM network, see the green and red boxes in Figs.~\ref{fig14}(d)-(e). The main reason is that the proposed ATFusion embeds the DIIM and ACIIM modules, which have a good ability to capture the discrepancy information and the common information of the source images.

\begin{table*}[!htbp]\scriptsize
    \caption{Quantitative assessment of the three different network structures. The best and the second best performance is marked by \textbf{bold} and \underline{underlined}, respectively.}
		\label{tab3}
		\begin{center}
			\begin{tabular}{ l l l l l l l l l l l l l}
				\hline
				Methods       &AG  &EN  &FMI &$Q_{abf}$ & SD  &SF   &SSIM &VIFF       \\		
                \hline
		w/o DIIM    &3.4038 &6.3734 &\textbf{0.9334} &0.6028 &8.1697 &\underline{0.0419} &0.9107 &\underline{1.0234} \\	
		w/o ACIIM   &\underline{3.4836} &\underline{6.5656} &0.9310 &\textbf{0.7067} &\underline{8.2095} &0.0411 &\textbf{0.9791} &0.9933 \\
		Ours        &\textbf{3.7619} &\textbf{6.5784} &\underline{0.9318} &\underline{0.7022} &\textbf{8.2459} &\textbf{0.0453} &\underline{0.9759} &\textbf{1.0370} \\
        \hline
            \end{tabular}
        \end{center}
        \label{w/oMoudle}
\end{table*}

Tab.~\ref{w/oMoudle} provides an objective comparison of our ATFusion with other two network structures. Our method can achieve the best performance for AG, EN, SD, SF, and VIFF, and the second-best performance for $Q_{abf}$, SSIM, and FMI. Therefore, the proposed DIIM and ACIIM modules can fully extract useful features, which enhance the retention of discrepancy information and the injection of common information in the fusion results.

\begin{figure*}[!t]
    \begin{center}
        \includegraphics[width=0.95\textwidth]{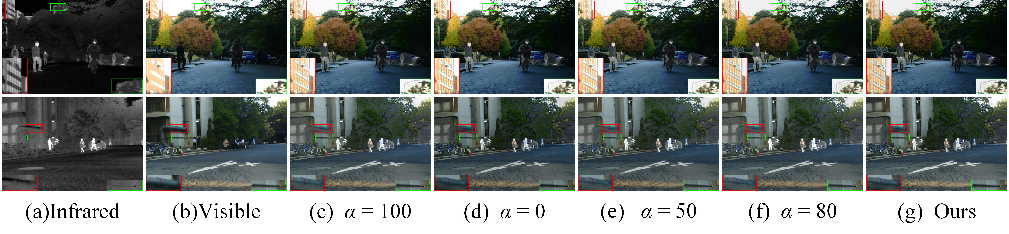}
    \end{center}
    	\caption{Fusion results obtained by the proposed method with five different pixel loss functions. (a) infrared image, (b) visible image, (c) the results of $\mathcal{L}_1$, (d) the results of $\mathcal{L}_2$, (e) the results of $\mathcal{L}_3$, (f) the results of $\mathcal{L}_4$, (g) Ours. }
        \label{fig15}
\end{figure*}

\emph{2) Ablation Study on the Pixel Loss Function}: To identify the impact of the segmented pixel loss function, we conduct four experiments by using different pixel loss functions: $\mathcal{L}_1$ ($\alpha = 100 $), $\mathcal{L}_2$ ($\alpha = 0 $), $\mathcal{L}_3$ ($\alpha = 50 $), and $\mathcal{L}_4$ ($\alpha = 80 $) in Eq.~(\ref{segmentseq}). $\mathcal{L}_1$ and $\mathcal{L}_2$ correspond to the pixel maximum and average constraints, respectively. Fig.~\ref{fig15} shows the fusion outcomes with the five different loss functions. From the green boxes in Figs.~\ref{fig15}(c),~(e), and~(f), it can be seen that the results generated by using $\mathcal{L}_1$, $\mathcal{L}_3$, and $\mathcal{L}_4$ have different degrees of detail loss. Besides, as the value of $\alpha$ increases, the degree of detail loss becomes more relevant. This is because the increasing value of $\alpha$ leads to more pixels in the image to be constrained by the maximum value selection loss, which brings to some information loss from one of the source images in fusion results. The results obtained by the $\mathcal{L}_2$ loss lack brightness information with respect to the results of the other loss functions, see the red box in Fig.~\ref{fig15}(d). This is because the core of the $\mathcal{L}_2$ loss is the pixel average constraint. Anyway, setting $\alpha$ to 20 results in a satisfactory trade-off between maintaining detail and preserving saliency.

\begin{table*}[!htbp]\scriptsize
    \caption{Quantitative assessment of the fusion results obtained by the proposed method with five different pixel loss functions. The best and the second best performance is marked by \textbf{bold} and \underline{underlined}, respectively.}
		\label{tab2}
		\begin{center}
			\begin{tabular}{ l l l l l l l l l l l l l}
				\hline
				Loss Functions       &AG  &EN  &FMI &$Q_{abf}$ & SD  &SF   &SSIM &VIFF\\
                \hline
		$\mathcal{L}_1$ ($\alpha = 100 $)   &3.6773 &\underline{6.5690} &\underline{0.9316} &0.7023 &\underline{8.2338} &\underline{0.0443} &0.9729 &1.0166 \\
		$\mathcal{L}_2$ ($\alpha = 0 $)     &\underline{3.7080} &6.5533 &0.9315 &0.7009 &8.2287 &0.0442 &\textbf{0.9774} &\textbf{1.0418} \\ 
		
 	$\mathcal{L}_3$ ($\alpha = 50 $)   &3.6805 &6.5671 &0.9315 &\underline{0.7034} &8.2197 &0.0441 &0.9750 &1.0169 \\
		$\mathcal{L}_4$ ($\alpha = 80 $) &3.6027 &6.5473 &0.9308 &\textbf{0.7042} &8.1918 &0.0432 &0.9731 &1.0131 \\
		Ours ($\alpha = 20 $)          &\textbf{3.7619} &\textbf{6.5784} &\textbf{0.9318} &0.7022 &\textbf{8.2459} &\textbf{0.0453} &\underline{0.9759} &\underline{1.0370} \\
        \hline
            \end{tabular}
        \end{center}
        \label{lossAblation}
\end{table*}

Tab.~\ref{lossAblation} reports the quantitative assessment of our ATFusion exploiting various loss functions. The average outcomes for the several evaluation metrics are tabulated in Tab.~\ref{lossAblation}. The best and second-best results for each metric are in boldface and underlined, respectively. It can be seen that the results generated by our method outperform the ones obtained by using the other loss functions. Therefore, setting $\alpha$ to 20 is a reasonable choice in our study.

\subsection{Analysis of Some Configurations}

\begin{figure*}[!t]
    \begin{center}
        \includegraphics[width=0.95\textwidth]{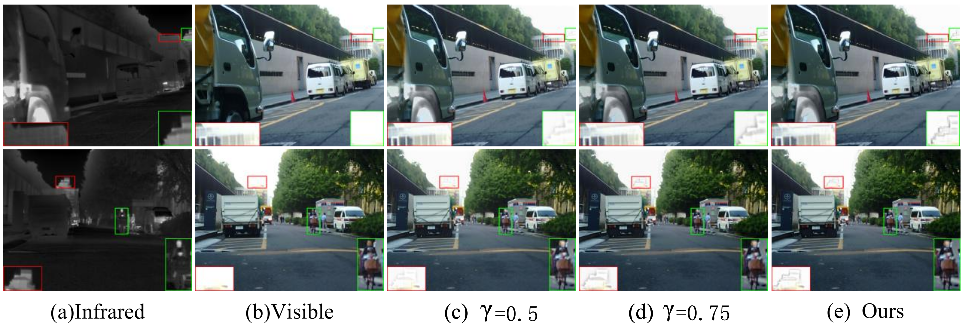}
    \end{center}
    	\caption{Fusion results obtained by the proposed method with different $\gamma$ values. (a) Infrared image, (b) visible image, (c)-(e) the qualitative results of different $\gamma$.}
        \label{fig16}
\end{figure*}

\begin{table*}[!htbp]\scriptsize
    \caption{Quantitative assessment of the fusion results obtained by different $\gamma$ values. The best and the second best performance is marked by \textbf{bold} and \underline{underlined}, respectively.}
		\label{gaMa}
		\begin{center}
			\begin{tabular}{ l l l l l l l l l l l l l}
				\hline
				Hyperparameter       &AG  &EN  &FMI &$Q_{abf}$ & SD  &SF   &SSIM &VIFF\\
                \hline
		$\gamma=0.5$   &3.6979 &\textbf{6.6916} &\underline{0.9316} &0.6918 &\textbf{8.3978} &0.0440 &\textbf{0.9772} &\underline{1.0281} \\
		$\gamma=0.75$  &\underline{3.7033} &6.5727 &0.9132 &\underline{0.7007} &\underline{8.2532} &\underline{0.0443} &0.9757 &1.0213 \\
		$\gamma=1.0$ (Ours)   &\textbf{3.7619} &\underline{6.5784} &\textbf{0.9318} &\textbf{0.7022} &8.2459 &\textbf{0.0453} &\underline{0.9759} &\textbf{1.0370} \\
        \hline
            \end{tabular}
        \end{center}
\end{table*}

\emph{1) Hyperparameter $\gamma$}: As mentioned in Section III.C, the hyperparameter $\gamma$ controls the weight of the pixel and texture losses in the loss function. Based on our experience, we set $\gamma$ to 0.5, 0.75, and 1.0. Fig.~\ref{fig16} displays the fusion images produced by ATFusion using different $\gamma$ values in the loss function. It can be seen that the output obtained with $\gamma = 1.0$ exhibits clear texture details, see the green box in Figs.~\ref{fig16}(c)-(e). Tab.~\ref{gaMa} presents an objective comparison of ATFusion using different $\gamma$ values. It is evident from the table that setting $\gamma$ to 1.0 yields the best performance.

\begin{table*}[!htbp]\scriptsize
    \caption{Quantitative assessment of fusion results obtained by the proposed method with different numbers of feature fusion modules. The best and the second best performance is marked by \textbf{bold} and \underline{underlined}, respectively.}
		\label{moudleNumber}
		\begin{center}
			\begin{tabular}{ l l l l l l l l l l l l l}
				\hline
				Number       &AG  &EN  &FMI &$Q_{abf}$ & SD  &SF   &SSIM &VIFF\\
                \hline
		Three   &\underline{3.7129} &\underline{6.4676} &\underline{0.9316} &\underline{0.6913} &\underline{8.1105} &\underline{0.0452} &0.9707 &\underline{1.0159} \\
		Two     &3.6630 &6.4268 &0.9304 &0.6900 &8.0113 &0.0444 &\underline{0.9710} &0.9844 \\
		One (Ours)  &\textbf{3.7619} &\textbf{6.5784} &\textbf{0.9318} &\textbf{0.7022} &\textbf{8.2459} &\textbf{0.0453} &\textbf{0.9759} &\textbf{1.0370} \\
        \hline
            \end{tabular}
        \end{center}
\end{table*}

\begin{figure*}[!t]
	\begin{center}
		\includegraphics[width=0.95\textwidth]{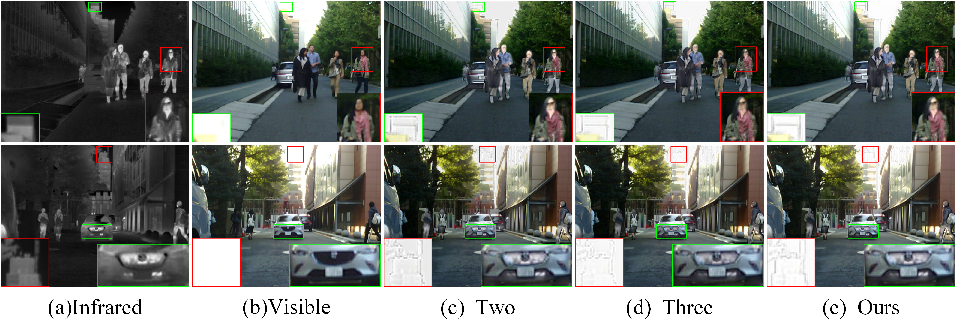}
	\end{center}
	\caption{Fusion results obtained by the proposed method with five different pixel loss functions. (a) Infrared image, (b) visible image, (c) the results of two feature fusion modules, (d) the results of three feature fusion modules, (e) Ours.}
	\label{fig17}
\end{figure*}

\emph{2) Numbers of Feature Fusion Modules}: We assembled one DIIM and two ACIIMs into a single feature fusion module, as shown in Fig.~\ref{framework}. The number of feature fusion modules can be decided by inspecting the fusion performance through objective metrics. Fig.~\ref{fig17} displays the fusion images produced by ATFusion using various numbers of feature fusion modules. It can be seen that using one feature fusion module in the proposed framework achieves the best overall effect, see the green and red boxes in Figs.~\ref{fig17}(c)-(e). Besides, Tab.~\ref{moudleNumber} reports the quantitative fusion results by using various numbers of feature fusion modules. We get the best results using one fusion model.

\subsection{Computational Efficiency}
The average computation time of nine different fusion methods across all IV datasets is listed in Tab.~\ref{time}. The most efficient and second most efficient methods are highlighted in bold and underlined, respectively. It can be seen that U2Fusion performs the highest efficiency while DenseFuse ranks second. The proposed ATFusion ranks third. This is because our method consumes time to alternately extract common information. Among all fusion methods, it is clear that our ATFusion is powerful and relatively lightweight in terms of fusion performance and running speed over others.

\begin{table*}[!htbp]\scriptsize
    \caption{Average running time of one image pair by using different fusion methods (unit: seconds). The best and the second best performance is marked by \textbf{bold} and \underline{underlined}, respectively.}
		\label{time}
		\begin{center}
			\begin{tabular}{ l l l l l l l l l l l l l}
				\hline
Methods &DenseFuse   &U2Fusion  &RFN-Nest &SwinFuse  &SwinFusion  &AEFusion  &DATFuse &LRRNet  &Ours \\		
                \hline
		Time (s) &\underline{0.0067} &\textbf{0.0056} &0.2562 &0.6888 &2.0278 &0.2945 &0.0463 &0.6052 &0.0080 \\		
        \hline
            \end{tabular}
        \end{center}
\end{table*}

\begin{figure*}[!t]
    \begin{center}
        \includegraphics[width=0.95\textwidth]{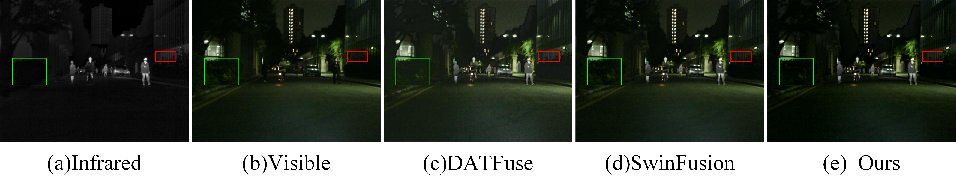}
    \end{center}
    	\caption{A failure case. Brightness degradation appears in fusion results.}
        \label{figConclusion}
\end{figure*}

\section{Conclusion}
In this paper, we observed that using the cross-attention mechanism for multimodal image fusion only extracted common information between the two source images, while ignoring the extraction of discrepancy information. Therefore, we modified cross-attention and proposed DIIM to extract discrepancy information from two modality images. Additionally, to mine and integrate long dependencies, we designed an alternate common information injection module (ACIIM) based on the vanilla cross-attention structure. Based on these two modules, we developed a novel VIF network named ATFusion. Specifically, we first applied the DIIM module to extract discrepancy information between source images, and then alternately extracted common information. To realize a good balance between salient information and detail information retention, we proposed a segmented pixel loss function, which utilized different constraint conditions for diverse pixel parts according to the importance of pixels. The proposed method has been validated for its effectiveness on the MSRS dataset. Furthermore, we applied this method to the RoadScene and RGB-NIR Scene datasets, indicating that our method has excellent generalization ability. In comparison with eight state-of-the-art fusion methods, extensive experiments show that ATFusion achieves the best fusion efficiency in both quality assessment and computation burden.

A limitation of our method is the poor ability to handle low-light scene information of the source images. Fig.~\ref{figConclusion} shows a representative case to illustrate this phenomenon. As can be seen, our method struggles to improve the quality degradation problem in night scenes. A possible reason for this visual performance is that our method tends to fuse the discrepancy information of the two images. However, when the source images all face dark conditions and are difficult to distinguish, our approach has difficulty in effectively extracting their discrepancy information. In future work, we will aim to design a low-light enhancement module to address the brightness degradation problem in VIF tasks.

\printcredits


\bibliographystyle{unsrt}
\bibliography{IPT_ATFusion}

\begin{thebibliography}{10}

\bibitem{yuan2024c}
Maoxun Yuan and Xingxing Wei.
\newblock C$^{2}$ former: Calibrated and complementary transformer for {RGB-Infrared} object detection.
\newblock {\em IEEE Transactions on Geoscience and Remote Sensing}, 2024.

\bibitem{mo2023robust}
Yan Mo, Xudong Kang, Shuo Zhang, Puhong Duan, and Shutao Li.
\newblock A robust infrared and visible image registration method for dual sensor {UAV} system.
\newblock {\em IEEE Transactions on Geoscience and Remote Sensing}, 2023.

\bibitem{zhang2022combined}
Jingwen Zhang, Xiaoxuan Zhou, Liyuan Li, Tingliang Hu, and Chen Fansheng.
\newblock A combined stripe noise removal and deblurring recovering method for thermal infrared remote sensing images.
\newblock {\em IEEE Transactions on Geoscience and Remote Sensing}, 60:1--14, 2022.

\bibitem{ma2019infrared}
Jiayi Ma, Yong Ma, and Chang Li.
\newblock Infrared and visible image fusion methods and applications: A survey.
\newblock {\em Information Fusion}, 45:153--178, 2019.

\bibitem{vivone2024deep}
Gemine Vivone, Liang-Jian Deng, Shangqi Deng, Danfeng Hong, Menghui Jiang, Chenyu Li, Wei Li, Huanfeng Shen, Xiao Wu, Jin-Liang Xiao, Jing Yao, Mengmeng Zhang, Jocelyn Chanussot, Salvador García, and Antonio Plaza.
\newblock Deep learning in remote sensing image fusion: Methods, protocols, data, and future perspectives.
\newblock {\em IEEE Geoscience and Remote Sensing Magazine}, 2024.

\bibitem{liu2018deep}
Yu~Liu, Xun Chen, Zengfu Wang, Z~Jane Wang, Rabab~K Ward, and Xuesong Wang.
\newblock Deep learning for pixel-level image fusion: Recent advances and future prospects.
\newblock {\em Information Fusion}, 42:158--173, 2018.

\bibitem{li2020mdlatlrr}
Hui Li, Xiao-Jun Wu, and Josef Kittler.
\newblock {MDLatLRR}: A novel decomposition method for infrared and visible image fusion.
\newblock {\em IEEE Transactions on Image Processing}, 29:4733--4746, 2020.

\bibitem{mou2013image}
Jiao Mou, Wei Gao, and Zongxi Song.
\newblock Image fusion based on non-negative matrix factorization and infrared feature extraction.
\newblock In {\em IEEE International Congress on Image and Signal Processing (CISP)}, volume~2, pages 1046--1050, 2013.

\bibitem{jian2021infrared}
Lihua Jian, Rakiba Rayhana, Ling Ma, Shaowu Wu, Zheng Liu, and Huiqin Jiang.
\newblock Infrared and visible image fusion based on deep decomposition network and saliency analysis.
\newblock {\em IEEE Transactions on Multimedia}, 24:3314--3326, 2021.

\bibitem{ma2016infrared}
Jiayi Ma, Chen Chen, Chang Li, and Jun Huang.
\newblock Infrared and visible image fusion via gradient transfer and total variation minimization.
\newblock {\em Information Fusion}, 31:100--109, 2016.

\bibitem{hou2019infrared}
Ruichao Hou, Rencan Nie, Dongming Zhou, Jinde Cao, and Dong Liu.
\newblock Infrared and visible images fusion using visual saliency and optimized spiking cortical model in non-subsampled shearlet transform domain.
\newblock {\em Multimedia Tools and Applications}, 78:28609--28632, 2019.

\bibitem{ren2021infrared}
Long Ren, Zhibin Pan, Jianzhong Cao, and Jiawen Liao.
\newblock Infrared and visible image fusion based on variational auto-encoder and infrared feature compensation.
\newblock {\em Infrared Physics \& Technology}, 117:103839, 2021.

\bibitem{xu2022multi}
Dongdong Xu, Ning Zhang, Yuxi Zhang, Zheng Li, Zhikang Zhao, and Yongcheng Wang.
\newblock Multi-scale unsupervised network for infrared and visible image fusion based on joint attention mechanism.
\newblock {\em Infrared Physics \& Technology}, 125:104242, 2022.

\bibitem{ding2021cmfa_net}
Zhaisheng Ding, Haiyan Li, Dongming Zhou, Hongsong Li, Yanyu Liu, and Ruichao Hou.
\newblock Cmfa\_net: A cross-modal feature aggregation network for infrared-visible image fusion.
\newblock {\em Infrared Physics \& Technology}, 118:103905, 2021.

\bibitem{yi2021dfpgan}
Shi Yi, Junjie Li, and Xuesong Yuan.
\newblock Dfpgan: Dual fusion path generative adversarial network for infrared and visible image fusion.
\newblock {\em Infrared Physics \& Technology}, 119:103947, 2021.

\bibitem{chen2021vision}
Xiangning Chen, Cho-Jui Hsieh, and Boqing Gong.
\newblock When vision transformers outperform resnets without pre-training or strong data augmentations.
\newblock {\em arXiv preprint arXiv:2106.01548}, 2021.

\bibitem{you2022hmf}
Tengfei You, Chanyue Wu, Yunpeng Bai, Dong Wang, Huibin Ge, and Ying Li.
\newblock {HMF-Former}: Spatio-spectral transformer for hyperspectral and multispectral image fusion.
\newblock {\em IEEE Geoscience and Remote Sensing Letters}, 20:1--5, 2022.

\bibitem{tang2023datfuse}
Wei Tang, Fazhi He, Yu~Liu, Yansong Duan, and Tongzhen Si.
\newblock {DATFuse}: Infrared and visible image fusion via dual attention transformer.
\newblock {\em IEEE Transactions on Circuits and Systems for Video Technology}, 2023.

\bibitem{ma2022swinfusion}
Jiayi Ma, Linfeng Tang, Fan Fan, Jun Huang, Xiaoguang Mei, and Yong Ma.
\newblock {SwinFusion}: Cross-domain long-range learning for general image fusion via swin transformer.
\newblock {\em IEEE/CAA Journal of Automatica Sinica}, 9(7):1200--1217, 2022.

\bibitem{chen2020enhanced}
Yuzhao Chen, Tao Dai, Xi~Xiao, Jian Lu, and Shu-Tao Xia.
\newblock Enhanced image restoration via supervised target feature transfer.
\newblock In {\em IEEE International Conference on Image Processing (ICIP)}, pages 1028--1032, 2020.

\bibitem{yuan2018deep}
Caihong Yuan, Chunyan Xu, Tianjiang Wang, Fang Liu, Zhiqiang Zhao, Ping Feng, and Jingjuan Guo.
\newblock Deep multi-instance learning for end-to-end person re-identification.
\newblock {\em Multimedia Tools and Applications}, 77:12437--12467, 2018.

\bibitem{ouyang2018video}
Deqiang Ouyang, Jie Shao, Yonghui Zhang, Yang Yang, and Heng~Tao Shen.
\newblock Video-based person re-identification via self-paced learning and deep reinforcement learning framework.
\newblock In {\em ACM International Conference on Multimedia}, pages 1562--1570, 2018.

\bibitem{zhang2018recent}
Yungang Zhang and Yu~Xiang.
\newblock Recent advances in deep learning for single image super-resolution.
\newblock In {\em International Conference on Brain Inspired Cognitive Systems}, pages 85--95, 2018.

\bibitem{liu2018infrared}
Yu~Liu, Xun Chen, Juan Cheng, Hu~Peng, and Zengfu Wang.
\newblock Infrared and visible image fusion with convolutional neural networks.
\newblock {\em International Journal of Wavelets, Multiresolution and Information Processing}, 16(03):1850018, 2018.

\bibitem{li2018densefuse}
Hui Li and Xiao-Jun Wu.
\newblock {DenseFuse}: A fusion approach to infrared and visible images.
\newblock {\em IEEE Transactions on Image Processing}, 28(5):2614--2623, 2018.

\bibitem{jian2020sedrfuse}
Lihua Jian, Xiaomin Yang, Zheng Liu, Gwanggil Jeon, Mingliang Gao, and David Chisholm.
\newblock {SEDRFuse}: A symmetric encoder--decoder with residual block network for infrared and visible image fusion.
\newblock {\em IEEE Transactions on Instrumentation and Measurement}, 70:1--15, 2020.

\bibitem{li2021rfn}
Hui Li, Xiao-Jun Wu, and Josef Kittler.
\newblock {RFN-Nest}: An end-to-end residual fusion network for infrared and visible images.
\newblock {\em Information Fusion}, 73:72--86, 2021.

\bibitem{ma2021stdfusionnet}
Jiayi Ma, Linfeng Tang, Meilong Xu, Hao Zhang, and Guobao Xiao.
\newblock {STDFusionNet}: An infrared and visible image fusion network based on salient target detection.
\newblock {\em IEEE Transactions on Instrumentation and Measurement}, 70:1--13, 2021.

\bibitem{xu2020u2fusion}
Han Xu, Jiayi Ma, Junjun Jiang, Xiaojie Guo, and Haibin Ling.
\newblock {U2Fusion}: A unified unsupervised image fusion network.
\newblock {\em IEEE Transactions on Pattern Analysis and Machine Intelligence}, 44(1):502--518, 2020.

\bibitem{wang2021unfusion}
Zhishe Wang, Junyao Wang, Yuanyuan Wu, Jiawei Xu, and Xiaoqin Zhang.
\newblock {UNFusion}: A unified multi-scale densely connected network for infrared and visible image fusion.
\newblock {\em IEEE Transactions on Circuits and Systems for Video Technology}, 32(6):3360--3374, 2021.

\bibitem{wang2022res2fusion}
Zhishe Wang, Yuanyuan Wu, Junyao Wang, Jiawei Xu, and Wenyu Shao.
\newblock {Res2Fusion}: Infrared and visible image fusion based on dense {Res2net} and double nonlocal attention models.
\newblock {\em IEEE Transactions on Instrumentation and Measurement}, 71:1--12, 2022.

\bibitem{ma2019fusiongan}
Jiayi Ma, Wei Yu, Pengwei Liang, Chang Li, and Junjun Jiang.
\newblock {FusionGAN}: A generative adversarial network for infrared and visible image fusion.
\newblock {\em Information fusion}, 48:11--26, 2019.

\bibitem{li2019coupled}
Qilei Li, Lu~Lu, Zhen Li, Wei Wu, Zheng Liu, Gwanggil Jeon, and Xiaomin Yang.
\newblock Coupled {GAN} with relativistic discriminators for infrared and visible images fusion.
\newblock {\em IEEE Sensors Journal}, 21(6):7458--7467, 2019.

\bibitem{ma2020ddcgan}
Jiayi Ma, Han Xu, Junjun Jiang, Xiaoguang Mei, and Xiao-Ping Zhang.
\newblock {DDcGAN}: A dual-discriminator conditional generative adversarial network for multi-resolution image fusion.
\newblock {\em IEEE Transactions on Image Processing}, 29:4980--4995, 2020.

\bibitem{ma2020ganmcc}
Jiayi Ma, Hao Zhang, Zhenfeng Shao, Pengwei Liang, and Han Xu.
\newblock {GANMcC}: A generative adversarial network with multiclassification constraints for infrared and visible image fusion.
\newblock {\em IEEE Transactions on Instrumentation and Measurement}, 70:1--14, 2020.

\bibitem{wang2023cross}
Zhishe Wang, Wenyu Shao, Yanlin Chen, Jiawei Xu, and Lei Zhang.
\newblock A cross-scale iterative attentional adversarial fusion network for infrared and visible images.
\newblock {\em IEEE Transactions on Circuits and Systems for Video Technology}, 33(8):3677--3688, 2023.

\bibitem{wang2022infrared}
Zhishe Wang, Wenyu Shao, Yanlin Chen, Jiawei Xu, and Xiaoqin Zhang.
\newblock Infrared and visible image fusion via interactive compensatory attention adversarial learning.
\newblock {\em IEEE Transactions on Multimedia}, 25:7800--7813, 2022.

\bibitem{vaswani2017attention}
Ashish Vaswani, Noam Shazeer, Niki Parmar, Jakob Uszkoreit, Llion Jones, Aidan~N Gomez, {\L}ukasz Kaiser, and Illia Polosukhin.
\newblock Attention is all you need.
\newblock {\em Advances in neural information processing systems}, 30, 2017.

\bibitem{dosovitskiy2020image}
Alexey Dosovitskiy, Lucas Beyer, Alexander Kolesnikov, Dirk Weissenborn, Xiaohua Zhai, Thomas Unterthiner, Mostafa Dehghani, Matthias Minderer, Georg Heigold, Sylvain Gelly, et~al.
\newblock An image is worth 16x16 words: Transformers for image recognition at scale.
\newblock {\em arXiv preprint arXiv:2010.11929}, 2020.

\bibitem{xie2022focal}
Tianming Xie, Zhonghao Zhang, Jing Tian, and Lihong Ma.
\newblock {Focal DETR}: Target-aware token design for transformer-based object detection.
\newblock {\em Sensors}, 22(22):8686, 2022.

\bibitem{lin2022infrared}
Jian Lin, Kai Zhang, Xi~Yang, Xiangzheng Cheng, and Chenhui Li.
\newblock Infrared dim and small target detection based on {U-Transformer}.
\newblock {\em Journal of Visual Communication and Image Representation}, 89:103684, 2022.

\bibitem{li2023feature}
Guanxiao Li, Ke~Zhang, Yu~Su, and Jingyu Wang.
\newblock Feature pre-inpainting enhanced transformer for video inpainting.
\newblock {\em Engineering Applications of Artificial Intelligence}, 123:106323, 2023.

\bibitem{gan2022hybrid}
Hongshan Gan and Yi~Wan.
\newblock A hybrid encoder transformer network for video inpainting.
\newblock In {\em International Conference on Computer, Control and Robotics (ICCCR)}, pages 230--234, 2022.

\bibitem{zou2023multi}
Beiji Zou, Zexin Ji, Chengzhang Zhu, Yulan Dai, Wensheng Zhang, and Xiaoyan Kui.
\newblock Multi-scale deformable transformer for multi-contrast knee {MRI} super-resolution.
\newblock {\em Biomedical Signal Processing and Control}, 79:104154, 2023.

\bibitem{ariav2023fully}
Ido Ariav and Israel Cohen.
\newblock Fully cross-attention transformer for guided depth super-resolution.
\newblock {\em Sensors}, 23(5):2723, 2023.

\bibitem{chen2023thfuse}
Jun Chen, Jianfeng Ding, Yang Yu, and Wenping Gong.
\newblock Thfuse: An infrared and visible image fusion network using transformer and hybrid feature extractor.
\newblock {\em Neurocomputing}, 527:71--82, 2023.

\bibitem{yu2023end}
Kaixin Yu, Xiaoming Yang, Seunggil Jeon, and Qingyu Dou.
\newblock An end-to-end medical image fusion network based on {Swin}-transformer.
\newblock {\em Microprocessors and Microsystems}, 98:104781, 2023.

\bibitem{vs2022image}
Vibashan Vs, Jeya Maria~Jose Valanarasu, Poojan Oza, and Vishal~M Patel.
\newblock Image fusion transformer.
\newblock In {\em IEEE International Conference on Image Processing (ICIP)}, pages 3566--3570, 2022.

\bibitem{wang2022swinfuse}
Zhishe Wang, Yanlin Chen, Wenyu Shao, Hui Li, and Lei Zhang.
\newblock {SwinFuse}: A residual swin transformer fusion network for infrared and visible images.
\newblock {\em IEEE Transactions on Instrumentation and Measurement}, 71:1--12, 2022.

\bibitem{xiao2021early}
Tete Xiao, Mannat Singh, Eric Mintun, Trevor Darrell, Piotr Doll{\'a}r, and Ross Girshick.
\newblock Early convolutions help transformers see better.
\newblock {\em Advances in Neural Information Processing Systems}, 34:30392--30400, 2021.

\bibitem{avenash2019semantic}
R~Avenash and Prashanth Viswanath.
\newblock Semantic segmentation of satellite images using a modified cnn with hard-swish activation function.
\newblock In {\em VISIGRAPP (4: VISAPP)}, pages 413--420, 2019.

\bibitem{2021CrossViT}
Chun~Fu Chen, Quanfu Fan, and Rameswar Panda.
\newblock {CrossViT}: Cross-attention multi-scale vision transformer for image classification.
\newblock {\em the IEEE/CVF International Conference on Computer Vision}, pages 357--366, 2021.

\bibitem{tang2022piafusion}
Linfeng Tang, Jiteng Yuan, Hao Zhang, Xingyu Jiang, and Jiayi Ma.
\newblock {PIAFusion}: A progressive infrared and visible image fusion network based on illumination aware.
\newblock {\em Information Fusion}, 83:79--92, 2022.

\bibitem{brown2011multi}
Matthew Brown and Sabine S{\"u}sstrunk.
\newblock Multi-spectral sift for scene category recognition.
\newblock In {\em CVPR 2011}, pages 177--184. IEEE.
\newblock 2011.

\bibitem{li2023lrrnet}
Hui Li, Tianyang Xu, Xiao-Jun Wu, Jiwen Lu, and Josef Kittler.
\newblock {Lrrnet}: A novel representation learning guided fusion network for infrared and visible images.
\newblock {\em IEEE Transactions on Pattern Analysis and Machine Intelligence}, 45:11040--11052, 2023.

\bibitem{li2023aefusion}
Bicao Li, Jiaxi Lu, Zhoufeng Liu, Zhuhong Shao, Chunlei Li, Yifan Du, and Jie Huang.
\newblock {AEFusion}: A multi-scale fusion network combining axial attention and entropy feature aggregation for infrared and visible images.
\newblock {\em Applied Soft Computing}, 132:109857, 2023.

\bibitem{liu2015general}
Yu~Liu, Shuping Liu, and Zengfu Wang.
\newblock A general framework for image fusion based on multi-scale transform and sparse representation.
\newblock {\em Information Fusion}, 24:147--164, 2015.

\bibitem{haghighat2011non}
Mohammad Bagher~Akbari Haghighat, Ali Aghagolzadeh, and Hadi Seyedarabi.
\newblock A non-reference image fusion metric based on mutual information of image features.
\newblock {\em Computers \& Electrical Engineering}, 37(5):744--756, 2011.

\bibitem{rao1997fibre}
Yun-Jiang Rao.
\newblock {In-fibre} bragg grating sensors.
\newblock {\em Measurement science and technology}, 8(4):355, 1997.

\bibitem{zheng2007new}
Yufeng Zheng, Edward~A Essock, Bruce~C Hansen, and Andrew~M Haun.
\newblock A new metric based on extended spatial frequency and its application to {DWT} based fusion algorithms.
\newblock {\em Information Fusion}, 8(2):177--192, 2007.

\bibitem{wang2004image}
Zhou Wang, Alan~C Bovik, Hamid~R Sheikh, and Eero~P Simoncelli.
\newblock Image quality assessment: from error visibility to structural similarity.
\newblock {\em IEEE Transactions on Image Processing}, 13(4):600--612, 2004.

\bibitem{han2013new}
Yu~Han, Yunze Cai, Yin Cao, and Xiaoming Xu.
\newblock A new image fusion performance metric based on visual information fidelity.
\newblock {\em Information Fusion}, 14(2):127--135, 2013.

\bibitem{piella2003new}
Gemma Piella and Henk Heijmans.
\newblock A new quality metric for image fusion.
\newblock In {\em IEEE International Conference on Image Processing}, volume~3, pages III--173, 2003.

\end{thebibliography}

\end{document}